\documentclass[aps,prl,twocolumn,superscriptaddress,nofootinbib]{revtex4-2}
\usepackage{amsmath,amssymb,graphicx,braket,tikz,hyperref,xcolor}
\usepackage[T1]{fontenc}
\usepackage{lmodern}
\usepackage[utf8]{inputenc}
\begin{document}

\title{Thermodynamic Signature of Logical Depth in Quantum Circuits}
\author{Issam Ibnouhsein}

\begin{abstract}
We demonstrate that the internal logical structure of a quantum circuit can leave a distinct thermodynamic signature under progressive decoherence. By comparing deep, conditionally branching circuits with shallow, uniform counterparts—while controlling for overall halting probability and physical resources—we show that branching architectures induce greater entropy flow into the environment. This effect is captured by a logical depth factor $L_d$, which quantifies entropy accumulation during environmental interactions. We validate our framework through detailed analysis of two 4-branch quantum circuits, demonstrating greater entropy production with $L_d \approx 1.615$ for conditional versus uniform architectures. An ancilla-based experimental protocol using controlled-phase gates provides a concrete pathway for detecting these thermodynamic signatures on current quantum platforms. Our results establish logical depth as a physically measurable quantity with implications for circuit design, compilation strategies, and verification protocols.
\end{abstract}

\maketitle

\section{Introduction}
\label{sec-intro}

The observable outputs of quantum circuits provide only partial information about their internal computational structure. Two circuits may yield the same overall measurement statistics while differing substantially in how information propagates through registers and how conditional operations unfold during execution. Such internal differences, though not captured by aggregate observables, reflect varying degrees of logical organization that are typically considered irrelevant from an input-output perspective.

Bennett's notion of \emph{logical depth}~\cite{Bennett1988} characterizes the sequential complexity of a computational process, measuring the number of dependent steps required to produce an output. In quantum circuits, logical depth acquires richer meaning: quantum superposition allows many computational paths to be traversed simultaneously, generating internal histories that maintain coherence during evolution and gradually fade through environmental interaction. 

Understanding the physical implications of such quantum logical structures requires examining their thermodynamic consequences. The thermodynamic cost of computation has been widely studied since Landauer's principle~\cite{Landauer1961} established the fundamental connection between information erasure and heat dissipation. Recent work in classical nanocomputing suggests that the structure of computation itself—independent of what is being computed—can influence thermodynamic behavior~\cite{Ercan2013}. For instance, studies on quantum-dot cellular automata have demonstrated that different clocking schemes, which control the sequence of information flow, lead to different heat dissipation profiles for circuits performing the same logical function~\cite{Ercan2014}. This raises a fundamental question: Can the structure of a quantum computation leave a detectable thermodynamic imprint?

In the quantum domain, generalizations of Landauer's principle and related frameworks have explored open-system dynamics~\cite{Breuer2002, Wolpert2019} and entropy accumulation during measurement~\cite{Deffner2016}, with experimental demonstrations now confirming these theoretical predictions~\cite{HernandezGomez2023, Clarke2024}. However, these analyses rarely isolate how internal logical structure contributes to such signatures. Existing approaches typically assign costs to specific state transformations~\cite{Faist2015,Sagawa2009} or non-unitary events~\cite{Reeb2014}, rather than examining how computational branching affects entropy production during unitary evolution.

In this work, we address this gap by showing that two quantum circuits, designed to have the same halting probability and equivalent physical resources, can exhibit distinct thermodynamic behavior due solely to their internal logical structure. We construct two carefully controlled architectures—one deep, with control-dependent branching, and one shallow, with uniform operations— and demonstrate that entropy flow under progressive decoherence differs between them. This effect is captured by a \emph{logical depth factor}~$L_d$, which quantifies the thermodynamic footprint of branching complexity. Our results establish logical depth as a physically meaningful quantity with implications for circuit optimization, compilation techniques, and verification protocols.

\section{Circuit Architectures}
\label{sec-framework}

We study two quantum circuit architectures that differ only in their internal logical structure. The systems employ three registers: a control register \(R_C\) with \(m\) qubits, a data register \(R_D\) with \(n\) qubits, and a single halting qubit \(Q_H\). To ensure a fair comparison, we carefully control for key variables except internal branching: both architectures use the same registers, time steps, and are tuned to have the same overall halting probability. All unitaries are drawn from a common ensemble—such as Haar-random or Clifford gates—with matched entangling power, and both circuits are compiled using the same universal gate set (e.g., \(\{\text{H}, \text{T}, \text{CNOT}\}\)) \cite{Barenco1995, Nielsen2010}. Crucially, we enforce branch-level complexity matching: each computational path within the deep circuit and the single path of the shallow circuit maintain identical gate count, depth, and entangling gate profiles. This standardization ensures that any observed entropy differences reflect logical structure rather than circuit complexity.

We model open-system dynamics via a memoryless interaction with an environment $E$, such as a Markovian dephasing bath, with coupling strength sufficient to create distinguishable environmental states, but weak enough to not disturb the main computation. In both architectures, the environment couples to the data register $R_D$ after each unitary transformation, encoding branch-dependent information into environmental degrees of freedom, while the halting qubit $Q_H$ remains isolated until the final measurement step.

Both circuits begin from the same initial superposition state:
\begin{equation}
\ket{\psi_{\text{init}}} = \sum_{i=0}^{2^m - 1} \alpha_i \ket{i}_{R_C} \otimes \ket{0}_{R_D} \otimes \ket{0}_{Q_H} \,.
\end{equation}

For the deep conditional architecture, computation proceeds through control-dependent operations. At each time step $t \in \{1, 2, \ldots, T\}$, we apply a controlled unitary:
\begin{equation}
U^{(t)}_{\text{deep}} = \sum_{i=0}^{2^m - 1} \ket{i}\bra{i}_{R_C} \otimes U_i^{(t)} \otimes I_{Q_H}\,,
\end{equation}
where the unitaries $U_i^{(t)}$ are drawn from a common ensemble. This creates distinct full evolutions $U_i = U_i^{(T)} \cdots U_i^{(1)}$ for each control value, implementing conditional branching. Each branch is constrained to have identical total gate complexity, ensuring that physical resources are matched across all computational paths while allowing logical operations to differ.

After each unitary transformation, the system interacts with the environment $E$:
\begin{equation}
\ket{\Psi_t} = \sum_{i=0}^{2^m - 1} \alpha_i \ket{i}_{R_C} \otimes \ket{\phi_i^{(t)}}_{R_D} \otimes \ket{0}_{Q_H} \otimes \ket{E_i^{(t)}} \,,
\end{equation}
where $\ket{\phi_i^{(t)}} = U_i^{(t)} \cdots U_i^{(1)} \ket{0}_{R_D}$ is the evolved data register state in branch $i$, and $\ket{E_i^{(t)}}$ is the corresponding environmental state.

Following the final computational step, a controlled halting operation is applied:
\begin{equation}
W = \sum_{i=0}^{2^m - 1} \ket{i}\bra{i}_{R_C} \otimes \left( \Pi_i \otimes X_{Q_H} + (I_{R_D} - \Pi_i) \otimes I_{Q_H} \right) \,,
\end{equation}
where $\Pi_i$ is a branch-specific halting projector. After this operation, the halting qubit couples to the environment for the first time, contributing to the final entropy production. The halting probability is determined by the projected amplitudes of the data register states:
\begin{equation}
P_{\text{halt}} = \sum_{i=0}^{2^m - 1} \lvert\alpha_i\rvert^2 \bra{\phi_i^{(T)}} \Pi_i \ket{\phi_i^{(T)}}\,.
\end{equation}

In contrast, the shallow architecture starts from the same initial state \(\ket{\psi_{\text{init}}}\) but applies global unitaries that act independently of the control register's state:
\begin{equation}
U_{\text{shallow}}^{(t)} = I_{R_C} \otimes U_{D}^{(t)}\,,
\end{equation}
where \(I_{R_C}\) is the identity on the control register, and the unitary \(U_{D}^{(t)}\) acting on the data register \(R_D\) is sampled from the same ensemble as the branch-specific unitaries \(U_i^{(t)}\) of the deep circuit. The sequence of unitaries on the data register is constructed such that the total gate complexity of the shallow path, summed over all time steps, matches the uniform total gate complexity of each individual deep branch. This design ensures comparability with the deep circuit while making the data register evolution identical across all computational branches, rendering them fundamentally indistinguishable to an environment coupled to \(R_D\).

After each time step \(t\), the system similarly undergoes uniform evolution and then interacts with the environment:
\begin{equation}
\ket{\Psi_t'} = \sum_{i=0}^{2^m - 1} \alpha_i \ket{i}_{R_C} \otimes \ket{\phi^{\prime(t)}}_{R_D} \otimes \ket{0}_{Q_H} \otimes \ket{E_i^{\prime(t)}}\,,
\end{equation}
where \(\ket{\phi^{\prime(t)}} = U_{\text{shallow}}^{(t)} \cdots U_{\text{shallow}}^{(1)} \ket{0}_{R_D}\) is independent of \(i\), and \(\ket{E_i^{\prime(t)}}\) denotes the corresponding environmental state entangled with branch \(i\).

The final data register state is:
\begin{equation}
\ket{\phi^{\prime(T)}} = U_{\text{shallow}}^{(T)} \cdots U_{\text{shallow}}^{(1)} \ket{0}_{R_D}\,.
\end{equation}

After applying the same halting operator $W$, the halting qubit couples to the environment for the first time, while the halting probability for the shallow circuit is given by:
\begin{equation}
P'_{\text{halt}} = \sum_{i=0}^{2^m - 1} |\alpha_i|^2 \bra{\phi^{\prime(T)}} \Pi_i \ket{\phi^{\prime(T)}}\,.
\end{equation}

To ensure the circuits are functionally equivalent from the perspective of a final measurement, we impose that the total halting probability of the shallow circuit matches that of the deep circuit:
\begin{equation}
\label{halt_constraint}
    \sum_{i=0}^{2^m-1} |\alpha_i|^2 \langle\phi'(T) | \Pi_i | \phi'(T)\rangle = \sum_{i=0}^{2^m-1} |\alpha_i|^2 \langle\phi_i(T) | \Pi_i | \phi_i(T)\rangle\,.
\end{equation}

This single global constraint can be satisfied by incorporating a steering operation in the final step, such as a general two-qubit unitary, to precisely adjust the halting probability. A two-qubit unitary is sufficient because it manipulates entangled correlations between any two data qubits, providing the necessary degrees of freedom to tune the aggregate halting probability to any target value. This operation adds only a small, constant overhead since a two-qubit unitary decomposes into at most 3 CNOTs and single-qubit rotations~\cite{Barenco1995}. To maintain comparable complexity, this overhead must be balanced within the shallow circuit's total resource budget, which is defined to match the uniform complexity of each deep branch. This can be achieved by either supplementing each deep branch with a similar final gate (treated as a fixed part of the branch-specific logic), or by reducing the shallow circuit's gate count in preceding steps.

The resulting architectures thus yield an identical overall outcome and use equivalent physical resources, differing solely in their internal logical structure. The deep circuit implements explicit conditional branching, whereas the shallow circuit performs flat, uniform operations. While the shallow circuit's fine-tuning requires information about the deep circuit's aggregate properties, this design-time informational complexity is distinct from the run-time thermodynamic cost that we investigate. This carefully constructed equivalence enables the isolation and measurement of the thermodynamic impact of logical organization.

\section{Entropy Flow and Logical Depth}
\label{sec-entropy-LD}

Having established our experimental framework, we now introduce the formalism for measuring thermodynamic signatures by tracking the entropy accumulated in the environment over time. Let $\rho_E^{(t)}$ and $\rho_E'^{(t)}$ be the reduced environmental states at time step $t$ for the deep and shallow circuits, respectively. The total entropy generated by each architecture, $\Delta S$, is the cumulative information leaked over the entire process. This is the sum of the von Neumann entropies, $S(\rho) = -\mathrm{Tr}(\rho \log \rho)$, from each of the $T$ computational steps, plus the entropy generated by the final halting operation.

We introduce the logical depth factor \(L_d\) as the ratio:
\begin{equation}
L_d = \frac{\Delta S_{\text{deep}}}{\Delta S_{\text{shallow}}}\,,
\end{equation}
which quantifies how internal branching structure amplifies entropy flow under decoherence. 

To model entropy production during computation in the two architectures, we consider environmental interaction through a dephasing Hamiltonian:
\begin{equation}
H_{\text{int}} = g \sum_{k=0}^{n-1} \hat{\sigma}_{z,k} \otimes B_k
\end{equation}
where $g$ is the coupling strength, $B_k$ are environmental operators, and $\hat{\sigma}_{z,k}$ are Pauli-Z operators monitoring the data register.

Under dephasing interaction, each branch generates a distinct environmental state:
\begin{equation}
|E_i^{(t)}\rangle = \exp\left(-ig \sum_{k=0}^{n-1} \langle\hat{\sigma}_{z,k}\rangle_{\phi_i^{(t)}} \int_0^t B_k(\tau) d\tau\right) |E_0\rangle\, ,
\end{equation}
where $|E_0\rangle$ denotes the initial state of the environment. 
The environmental states overlap can be approximated by:
\begin{equation}
|\langle E_i^{(t)} | E_j^{(t)} \rangle|^2 \approx \exp(-\gamma \cdot D_{ij}^{(t)})
\end{equation}
where $\gamma$ is an effective dephasing strength parameter that encapsulates the fundamental coupling $g$ and the time-integrated correlation functions of the environmental operators $B_k$, and $D_{ij}^{(t)}$ is the branch distinguishability parameter:
\begin{equation}
D_{ij}^{(t)} = \sum_{k=0}^{n-1} \left(\langle\phi_i^{(t)}|\hat{\sigma}_{z,k}|\phi_i^{(t)}\rangle - \langle\phi_j^{(t)}|\hat{\sigma}_{z,k}|\phi_j^{(t)}\rangle\right)^2
\end{equation}
which measures how different the environmental signatures of branches $i$ and $j$ are at time $t$. For any environment with $\gamma > 0$, environmental overlap decreases with the circuit's intrinsic branch distinguishability $D_{ij}^{(t)}$. The minimal coupling strength required for measurable signatures is determined by these distinguishability values, with a threshold condition: 
\begin{equation}
\label{obs-cond}
 \gamma \gtrsim 1/\min_{\{i,j\}} \sum_{t=1}^T D_{ij}^{(t)}   
\end{equation}
Whether these signature are detectable in practice depends on the experimental platform's precision and noise levels.

This distinguishability formalism reveals fundamental differences between architectures that share identical physical resources and halting probabilities, but differ in their internal branching logic. In the shallow circuit, uniform evolution ensures that the distinguishability parameters satisfy $D_{ij}'^{(t)} = 0$ for all branch pairs, corresponding to identical environmental states $\ket{E_i'^{(t)}}$ across branches:
\begin{equation}
|\braket{E_i'^{(t)} | E_j'^{(t)}}| = 1 \quad \text{for } i \neq j\,.
\end{equation}
The reduced environmental state at time $t$, 
\begin{equation}
\rho_E'^{(t)} = \sum_{i,j=0}^{2^m - 1} \alpha_i \alpha_j^* \ket{E_i'^{(t)}} \bra{E_j'^{(t)}}\,,
\end{equation}
therefore remains pure. This indicates that no which-path information is acquired and entanglement occurs only with the data register, limiting environmental information to only the $n$ data qubits. As a result, the entropy per time step remains low, typically bounded by:
\begin{equation}
S(\rho_E'^{(t)}) \lesssim n \log 2\,.
\end{equation}

Since the halting qubit stays in a fixed pure state throughout the computation and only becomes entangled with the environment during the final halting operation, the total entropy accumulated over $T$ computational steps plus the final halting operation is bounded by:
\begin{equation}
\label{delta_shallow}
\Delta S_{\text{shallow}} \lesssim (T \times n + 1) \, \log 2\,.
\end{equation}

In contrast, the deep architecture applies control-dependent operations: each control value \( i \) generates a distinct evolution sequence \( |\phi_i^{(t)}\rangle = U_i^{(t)} \cdots U_i^{(1)} |0\rangle_{R_D} \). For typical unitary families where \( U_i^{(t)} \neq U_j^{(t)} \) for many branch pairs \( i, j \), the resulting data register states diverge significantly across branches, leading to non-zero distinguishability parameters \( D_{ij}^{(t)} > 0 \) for $i \neq j$. This causes the corresponding environmental states to become nearly orthogonal:
\begin{equation}
|\braket{E_i^{(t)} | E_j^{(t)}}| \ll 1 \quad \text{for } i \neq j\,,
\end{equation}
which in turn yields a reduced environmental state that is approximately a diagonal mixture:
\begin{equation}
\rho_E^{(t)} \approx \sum_{i=0}^{2^m - 1} |\alpha_i|^2 \ket{E_i^{(t)}} \bra{E_i^{(t)}}\,.
\end{equation}

Unlike the shallow architecture—where all branches evolve identically—the deep circuit renders the control register \emph{indirectly} observable to the environment through branch-dependent data evolution. This effectively gives the environment access to all \( (m + n) \) qubits of information, encoding up to \( m \) bits of which-path information in addition to the \( n \) bits arising from data-register entanglement. The entropy generated per step is typically bounded by:
\begin{equation}
S(\rho_E^{(t)}) \lesssim (m+n) \, \log 2\,.
\end{equation}
Accordingly, the total entropy accumulated over $T$ time steps plus the final halting operation satisfies:
\begin{equation}
\label{delta_deep}
\Delta S_{\text{deep}} \lesssim (T \times (m+n) + 1) \, \log 2\,.
\end{equation}

From inequalities~\eqref{delta_shallow} and~\eqref{delta_deep}, and assuming the entropy bounds for shallow and deep circuits are approximately saturated, the resulting logical depth factor satisfies:
\begin{equation}
L_d \approx 1 + \frac{m}{n + \frac{1}{T}}\, .
\end{equation}
Such saturation conditions can be achieved through careful design—using sufficiently diverse control unitaries, well-distributed control amplitudes, and balanced entanglement with the data register—combined with coupling strength sufficient to resolve branch distinguishability without destroying the computational logic.

The logical depth factor increases primarily with the number of control qubits \(m\), which determines the number of distinct computational branches. This scaling remains valid as long as entanglement with the data register does not saturate the environment's capacity to record distinguishable information, and the decoherence time \(T\) is not too short. Although entropy accumulates over time, the growth of \(L_d\) eventually plateaus once the environment has fully resolved the accessible branching structure. The specific values of entropy bounds depend on coupling strength, but the fundamental ratio $L_d$ reflects the underlying information-theoretic difference between architectures.

This thermodynamic distinction between deep and shallow circuits thus constitutes a fundamental signature of internal computational structure. Under progressive decoherence, the environment responds not only to the final measurement outcomes but also to the structure by which they are computed, resulting in measurable signatures of logical depth.

\section{Example Circuits Analysis}
\label{sec-example}

To illustrate our framework, we analyze and compare two specific quantum circuit configurations—one deep (branching) and one shallow (uniform)—with concrete parameter values. Both circuits operate on a quantum system with a control register $R_C$ containing $m = 2$ qubits that generates $2^m = 4$ computational branches, a data register $R_D$ with $n = 3$ qubits ordered as $|q_0 q_1 q_2\rangle$, and a single halting qubit $Q_H$. The computation evolves over $T = 4$ time steps. Both circuits use the same initial state with uniform amplitudes across all control branches:
\begin{equation}
|\psi_{\text{init}}\rangle = \frac{1}{2}(|00\rangle + |01\rangle + |10\rangle + |11\rangle)_{R_C} \otimes |000\rangle_{R_D} \otimes |0\rangle_{Q_H}
\end{equation}
For simplicity, the halting condition uses a uniform projector $\Pi_{\text{halt}} = |0\rangle\langle 0|_2$ on $q_2$. After time step $T$, a controlled-X operation is applied to the halting qubit $Q_H$ if the data register state meets the halting condition.

For our specific example, the environmental monitoring operators are:
\begin{align}
\hat{\sigma}_{z,0} = Z_0\,,\quad \hat{\sigma}_{z,1} = Z_1 \,,\quad \hat{\sigma}_{z,2} = Z_2
\end{align}
The halting qubit $Q_H$ remains isolated from environmental interaction during computation, contributing entropy only when measured at the final step.

For the deep circuit, each control state $i \in \{0,1,2,3\}$, corresponding to the computational basis states $\{|00\rangle, |01\rangle, |10\rangle, |11\rangle\}$, activates a distinct sequence of unitary operations:

\begin{align}
\text{Branch 0}: \quad &U_0^{(1)} = H_0\,, \quad U_0^{(2)} = X_1\,, \nonumber\\
&U_0^{(3)} = \text{CNOT}_{0,2}\,, \quad U_0^{(4)} = R_Z(\pi/4)_1 \nonumber\\[0.5em]
\text{Branch 1}: \quad &U_1^{(1)} = H_1\,, \quad U_1^{(2)} = X_0\,, \nonumber\\
&U_1^{(3)} = \text{CNOT}_{1,2}\,, \quad U_1^{(4)} = R_Z(\pi/3)_0 \nonumber\\[0.5em]
\text{Branch 2}: \quad &U_2^{(1)} = H_2\,, \quad U_2^{(2)} = X_0\,, \nonumber\\
&U_2^{(3)} = \text{CNOT}_{0,1}\,, \quad U_2^{(4)} = R_Z(\pi/2)_2 \nonumber\\[0.5em]
\text{Branch 3}: \quad &U_3^{(1)} = H_0\,, \quad U_3^{(2)} = X_2\,, \nonumber\\
&U_3^{(3)} = \text{CNOT}_{1,0}\,, \quad U_3^{(4)} = R_Z(\pi/6)_1
\end{align}

To illustrate the system evolution, we follow Branch 0 starting from $|\phi_0^{(0)}\rangle = |000\rangle$. Application of $U_0^{(1)} = H_0$ yields:

\begin{equation}
|\phi_0^{(1)}\rangle = \frac{1}{\sqrt{2}}(|000\rangle + |100\rangle)
\end{equation}
The expectation values of the environmental monitoring operators are $(\langle\hat{\sigma}_{z,0}\rangle, \langle\hat{\sigma}_{z,1}\rangle, \langle\hat{\sigma}_{z,2}\rangle) = (0, 1, 1)$. Continuing the evolution, $U_0^{(2)} = X_1$ produces:
\begin{equation}
|\phi_0^{(2)}\rangle = \frac{1}{\sqrt{2}}(|010\rangle + |110\rangle)
\end{equation}
with environmental monitoring expectation values $(0, -1, 1)$. The third step applies $U_0^{(3)} = \text{CNOT}_{0,2}$ to give:
\begin{equation}
|\phi_0^{(3)}\rangle = \frac{1}{\sqrt{2}}(|010\rangle + |111\rangle)
\end{equation}
with environmental monitoring expectation values $(0, -1, 0)$. Finally, $U_0^{(4)} = R_Z(\pi/4)_1$ produces the Branch 0 final state:
\begin{equation}
|\phi_0^{(4)}\rangle = \frac{e^{i\pi/8}}{\sqrt{2}} \left( |010\rangle + |111\rangle \right)
\end{equation}
with final environmental monitoring expectation values $(0, -1, 0)$.

The remaining branches yield distinct final states:
\begin{align}
|\phi_1^{(4)}\rangle &= \frac{e^{i\pi/6}}{\sqrt{2}}\left(|100\rangle + |111\rangle\right)\,, \nonumber\\
|\phi_2^{(4)}\rangle &= \frac{1}{\sqrt{2}}\left(e^{-i\pi/4}|110\rangle + e^{i\pi/4}|111\rangle\right)\,, \nonumber\\
|\phi_3^{(4)}\rangle &= \frac{e^{-i\pi/12}}{\sqrt{2}}\left(|001\rangle + |101\rangle\right)
\end{align}
with final environmental monitoring expectation values of $(-1, 0, 0)$ for Branch 1, $(-1, -1, 0)$ for Branch 2, and $(0, 1, -1)$ for Branch 3.

The shallow circuit, in contrast, applies uniform operations across all branches:
\begin{align}
U_{\text{shallow}}^{(1)} &= H_0\,, \quad U_{\text{shallow}}^{(2)} = X_1\,, \nonumber\\
U_{\text{shallow}}^{(3)} &= \text{CNOT}_{0,1}\,, \quad U_{\text{shallow}}^{(4)} = R_Y(\theta)_2
\end{align}

Evolution proceeds identically for all control states through the first three time steps, yielding the state $|\phi'^{(3)}\rangle = \frac{1}{\sqrt{2}}(|010\rangle + |100\rangle)$ with environmental monitoring expectation values $(0, 0, 1)$. The $R_Y(\theta)$ rotation on $q_2$ produces an identical vector across all paths:
\begin{equation}
\begin{aligned}
|\phi'^{(4)}\rangle &= \frac{1}{\sqrt{2}}\bigg[ \cos\left(\frac{\theta}{2}\right)\left(|010\rangle + |100\rangle\right) \\
&\quad + \sin\left(\frac{\theta}{2}\right)\left(|011\rangle + |101\rangle\right) \bigg]
\end{aligned}
\end{equation}
with final environmental monitoring expectation values of $(0, 0, \cos(\theta))$.

The example circuits are deliberately designed to reflect the principle of uniform complexity established in our theoretical framework. Each of the four deep branches, as well as the single shallow path, is constructed with the same gate count and complexity profile: three single-qubit gates and one two-qubit CNOT gate. Although the gate order is identical in this specific example, this is not required by the general framework. This strict control over physical resources ensures zero complexity variance across computational paths, enabling a clean isolation of the thermodynamic effects of logical structure.

Functional equivalence between the two architectures is enforced through the appropriate choice of the rotation angle \( \theta \). For the deep circuit, the halting probabilities under the projector \( \Pi_{\text{halt}} = |0\rangle\langle 0|_2 \) on qubit \( q_2 \) are \( P_{\text{halt}}^{(0)} = 0.5 \), \( P_{\text{halt}}^{(1)} = 0.5 \), \( P_{\text{halt}}^{(2)} = 0.5 \), and \( P_{\text{halt}}^{(3)} = 0.0 \), yielding an average halting probability of \(P_{\text{halt}} = 0.375 \). The halting probability of the shallow circuit is given by \( P'_{\text{halt}} = \cos^2(\theta/2) \). Setting \( \cos^2(\theta/2) = 0.375 \) yields \( \theta = 1.824\ \mathrm{rad} \), ensuring the two circuits exhibit matched overall statistical behavior.

Applying the distinguishability formalism from Section~\ref{sec-entropy-LD} to our specific circuits, we compute the branch distinguishability parameters \( D_{ij}^{(t)} \) using the previously computed expectations values of the environmental monitoring operators. Table~\ref{dists_table} reports both the final-step and cumulative distinguishability values for the deep circuit, showing that entropy is generated progressively over time rather than solely at the final measurement. In contrast, the shallow circuit maintains \( D_{ij}'^{(t)} = 0 \) for all pairs and times, confirming perfect branch indistinguishability and strictly limiting the environment's information access to the data register.

\begin{table}[h]
\centering
\setlength{\tabcolsep}{12pt}
\renewcommand{\arraystretch}{1.4}
\begin{tabular}{ccc}
\hline\hline
Branch Pair & Final Step & Accumulated \\
\hline
$(0,1)$ & 2.0 & 8.0 \\
$(0,2)$ & 1.0 & 10.0 \\
$(0,3)$ & 5.0 & 18.0 \\
$(1,2)$ & 1.0 & 6.0 \\
$(1,3)$ & 3.0 & 14.0 \\
$(2,3)$ & 6.0 & 16.0 \\
\hline\hline
\end{tabular}
\caption{Branch distinguishability measures for the deep circuit showing final-step values $D_{ij}^{(4)}$ and accumulated values $\sum_{t=1}^4 D_{ij}^{(t)}$ over all time steps.}
\label{dists_table}
\end{table}

This difference in branch distinguishability has direct thermodynamic implications. For our specific parameters, the deep circuit provides the environment with access to all \( m + n = 5 \) qubits of information, while the shallow circuit exposes only the \( n = 3 \) data qubits. When the environmental coupling is strong enough to resolve these differences (\( \gamma \gtrsim 1/6.0 \approx 0.17 \), based on Eq.\ref{obs-cond} and the minimum accumulated distinguishability from Table~\ref{dists_table}) this disparity in accessible information directly determines the entropy production bounds:
\begin{equation}
\begin{aligned}
\Delta S_{\text{deep}} &= (4 \times (2+3)+1) \times \log2 = 21\,\log2  \\
\Delta S_{\text{shallow}} &= (4 \times 3+1) \times \log2 = 13\,\log2
\end{aligned}
\end{equation}
The logical depth factor in this limit is:
\begin{equation}
L_d = \frac{\Delta S_{\text{deep}}}{\Delta S_{\text{shallow}}} = \frac{21}{13} \approx 1.615
\end{equation}
Thus, the deep conditional architecture has the capacity to generate up to $61.5\%$ more entropy than the shallow uniform architecture. This illustrates the physical mechanism whereby logical depth amplifies entropy production through environmental distinguishability of computational branches.

\section{Thermodynamic Witness}
\label{sec-witness}

Our analysis reveals that quantum logical structure leaves a measurable thermodynamic imprint: the environment records not only \emph{what} is computed, but also \emph{how} it is computed, through its sensitivity to the branching structure of the computation.

Experimentally verifying this thermodynamic signature can be achieved by engineering a controlled ``toy environment'' using a single high-fidelity ancilla qubit. Rather than relying on uncontrolled environmental decoherence, we simulate environmental monitoring by weakly coupling this auxiliary qubit to the data register \( R_D \) via controlled-phase gates, C-PHASE(\( \phi \)), applied from each data qubit to the ancilla after every unitary operation. The rotation angle \( \phi \) serves as a tunable coupling strength, and the ancilla is never reset between steps, allowing it to coherently accumulate phase information that encodes the branch-dependent computational trajectories.

Observability of this thermodynamic imprint depends on achieving sufficient phase separation between ancilla states associated with different computational branches. Adapting the environmental overlap formalism from Section~\ref{sec-entropy-LD}, with the ancilla parameter \( \phi^2 \) replacing the bath coupling \( \gamma \), the condition becomes:
\begin{equation}
\phi^2 \sum_{t=1}^T D_{ij}^{(t)} \gtrsim 1\,.
\end{equation}

For the shallow circuit, uniform evolution preserves the ancilla in a pure state with zero von Neumann entropy, while the deep circuit's branch-dependent evolution creates an incoherent mixture:
\begin{equation}
\rho_{\text{anc}} \approx \sum_i |\alpha_i|^2 |\text{anc}_i\rangle\langle\text{anc}_i|\, .
\end{equation}
The key experimental observable is the ancilla's purity, \( \text{Tr}(\rho_{\text{anc}}^2) \), which equals 1 for a pure state and drops below 1 when decoherence reveals internal logical structure. In this setting, the shallow architecture yields \( P_{\text{shallow}} = 1 \), while the deep architecture gives \( P_{\text{deep}} < 1 \), providing a clear experimental signature of branching logic. This protocol, readily accessible via single-qubit quantum state tomography~\cite{Nielsen2010}, serves as a simplified contrast witness: unlike the theoretical framework, which tracks absolute entropy with non-zero baselines, the ancilla detects only relative differences across computational paths.

In our four-branch circuit example from Section \ref{sec-example}, the minimal accumulated distinguishability of 6.0 requires \( \phi \gtrsim 0.408 \) rad. We choose \( \phi = 0.5 \) rad, providing a safety margin while remaining in the weak-coupling regime. For this coupling strength and circuit pair, we predict \( P_{\text{shallow}} = 1.0 \) and \( P_{\text{deep}} \approx 0.797 \), corresponding to a significant purity drop of over 20\% in the deep architecture. This signal is well within the reach of current superconducting quantum platforms with gate fidelities exceeding 99\%~\cite{Krantz2019,GoogleWillow2024}, demonstrating the practical feasibility of measuring logical depth signatures.

\section{Practical Implications}
\label{sec-implications}

With both the theoretical framework and experimental feasibility of thermodynamic witnesses to logical depth now demonstrated, we examine the broader implications of these findings.

First, these entropy signatures suggest new principles for decoherence-aware circuit optimization. While standard techniques prioritize minimizing gate count or depth~\cite{Preskill2018}, our results highlight the benefit of reducing branching complexity to limit entropy flow. Flattening conditional logic—even at the cost of increased gate count—can enhance coherence in noisy devices, a strategy particularly relevant for fault-tolerant quantum computing~\cite{GoogleWillow2024}.

Second, entropy flow imposes a thermodynamic constraint on quantum circuit compilation. While this process typically preserves observable behavior, it can reshape internal logical depth. Reducing branching during the compilation process can lower entropy production, enhancing robustness. This makes internal circuit structure a key design consideration, complementing standard metrics like circuit size and fidelity—particularly in near-term quantum devices where decoherence remains a central challenge~\cite{Preskill2018}.

Third, thermodynamic signatures offer new tools for circuit verification. Even when two circuits yield identical halting probabilities, their ancilla purity signatures can differ if their internal logic diverges. This enables verification protocols that confirm a specific logical structure—not merely functional output—was realized during execution, which is critical for secure delegated computing or quantum cloud services.

Finally, these results deepen the connection between computation and thermodynamics. The logical depth factor \(L_d\) quantifies the entropy cost of traversing complex computational paths~\cite{Faist2015,Deffner2019}, showing that entropy flow reflects not only final-state measurement or erasure, but also the internal structure of unitary evolution itself. This includes the impact of unrealized branches in superposition, whose mere distinguishability contributes to entropy generation. 

This view reframes decoherence not just as noise, but as a physical recording of internal logic, aligning with the framework of Quantum Darwinism~\cite{Zurek2009}. The control register basis states $\{|i\rangle_{R_C}\}$ act as computational \emph{pointer states} that the environment repeatedly interrogates at each time step. For the deep circuit, this process creates a \emph{redundant encoding} of which-path information, as multiple environmental fragments independently record the same branch index $i$, while the shallow circuit produces no such record since its evolution remains blind to the pointer states. The increased entropy flow in the deep case reflects this richer internal structure. Logical organization thus has concrete thermodynamic implications, extending beyond abstract complexity to physically measurable consequences.

\section{Limitations and Assumptions}
\label{sec-limitations}

Our theoretical and experimental framework relies on simplifying assumptions that delimit the generality of our conclusions. Most importantly, the entropy-based signatures of logical depth require a coupling regime where the environment is sensitive to internal computational structure without collapsing the superposition prematurely. Strong decoherence would destroy the very correlations our protocol aims to probe, while weak coupling may fail to resolve branch differences, limiting observability.

We also model the environment as memoryless, assuming independent interactions after each time step. While this assumption makes the analysis simpler, real environments often exhibit non-Markovian behavior. Such memory effects could create misleading correlations that affect the results.

Moreover, our protocol presumes high-fidelity quantum operations and well-calibrated environmental interactions. In practice, gate errors and imperfect initialization could blur the entropy contrast between architectures, especially in deeper or noisier circuits. While our example circuit demonstrates that observable purity drops are already within reach of current superconducting platforms, fine-grained control over interaction strength remains a technical challenge, especially when operating close to the observability threshold.

Finally, while our theoretical framework applies to circuits of arbitrary size, experimental implementation remains more feasible in small to intermediate-scale systems, where entropy signatures remain detectable before errors accumulate. Scaling the protocol to larger circuits would require robust error mitigation and precise control over environmental interactions.

\section{Conclusion}
\label{sec-conclusion}

We have demonstrated that quantum circuits with matched halting probabilities and identical physical resources can nonetheless display distinct thermodynamic behavior due to differences in computational logic. By controlling for qubit registers, time steps, unitary ensembles, gate count, and entangling power—while isolating logical structure as the only varying factor—we have shown how branching complexity can leave a measurable thermodynamic trace.

This trace is captured by the logical depth factor $L_d$, defined as the ratio of cumulative environmental entropy generated by deep versus shallow architectures. Our concrete circuit example illustrates this effect quantitatively, while our ancilla-based witness protocol demonstrates that these entropy differences are experimentally accessible with coupling strengths readily achievable in current quantum platforms. Our results establish that logical depth, interpreted as conditional path complexity, is not merely an abstract syntactic quantity but a physically meaningful and experimentally accessible measure of computational structure.

These findings offer several practical applications. The framework enables new approaches to quantum circuit design and verification, providing path-sensitive diagnostics that capture structural differences missed by conventional benchmarks. Our analysis also reveals a fundamental thermodynamic trade-off: branching logic enhances algorithmic expressivity but increases entropy flow, reducing coherence in open-system settings. This suggests compilation strategies that flatten conditional logic to yield cleaner implementations in noise-constrained architectures.

More broadly, our results deepen the established connection between quantum computation and thermodynamics, demonstrating that logical depth governs both the informational complexity of quantum algorithms and their physical footprint during execution. This perspective supports a unified view where internal logical structure, environmental interaction, and thermodynamic cost emerge as inseparable aspects of quantum computation.

Future work could extend this framework in several directions. Developing robust entropy signatures that remain detectable under stronger environmental coupling and non-Markovian conditions would broaden experimental applicability. Exploring thermodynamic witnesses in fault-tolerant architectures, measurement-based quantum computing, and hybrid quantum-classical systems could reveal how logical depth signatures emerge across different computational paradigms. Additionally, investigating connections to other complexity measures may deepen our understanding of the fundamental thermodynamic limits of quantum information processing.

\bibliographystyle{unsrt}
\bibliography{refs}

\end{document}